\documentclass[prd,preprintnumbers,amsmath,amssymb,superscriptaddress,nofootinbib]{revtex4}

\usepackage{graphicx}
\usepackage{dcolumn}
\usepackage{bm}
\usepackage{xcolor}
\usepackage{caption}
\usepackage{slashed}
\usepackage{hyperref}

\newcommand{\nn}{\nonumber}

\def\beq{\begin{equation}}
\def\eeq{\end{equation}}
\def\bqa{\begin{eqnarray}}
\def\eqa{\end{eqnarray}}

\def\dfrac{\displaystyle\frac}

\allowdisplaybreaks[2]

\begin{document}

\preprint{JLAB-THY-24-4020}

\title{Soft pattern of gravitational Rutherford scattering from heavy target mass expansion}

\author{Yu Jia~\footnote{jiay@ihep.ac.cn}}
\affiliation{Institute of High Energy Physics, Chinese Academy of Sciences, Beijing 100049, China\vspace{0.2cm}}
\affiliation{School of Physics, University of Chinese Academy of Sciences, Beijing 100049, China\vspace{0.2cm}}

\author{Jichen Pan~\footnote{panjc@ihep.ac.cn}}
\affiliation{Institute of High Energy Physics, Chinese Academy of Sciences, Beijing 100049, China\vspace{0.2cm}}
\affiliation{School of Physics, University of Chinese Academy of Sciences, Beijing 100049, China\vspace{0.2cm}}

\author{Jia-Yue Zhang~\footnote{jzhang@jlab.org}}
\affiliation{Theory Center, Jefferson Lab, Newport News, Virginia 23606, USA\vspace{0.2cm}}
\affiliation{Institute of High Energy Physics, Chinese Academy of Sciences, Beijing 100049, China\vspace{0.2cm}}

\date{\today}

\begin{abstract}
We investigate the soft behavior of the tree-level Rutherford scattering processes mediated via $t$-channel  one-graviton exchange.
We consider two types of Rutherford scattering processes, {\it e.g.}, a low-energy massless structureless projectile (up to spin-$1$) hits
a static massive composite particle carrying various spins (up to spin-$2$), and a slowly-moving light projectile hits a heavy static composite target.
The unpolarized cross sections in the first type are found to exhibit universal forms at the first two orders in $1/M$ expansion,
yet differ at the next-to-next-to-leading order, though some terms at this order
still remain universal or depend on the target spin in a definite manner.
The unpolarized cross sections in the second type are universal at the lowest order in projectile velocity expansion and through all orders in $1/M$,
independent of the spins of both projectile and target. The universality partially breaks down at relative order-$v^2/M^2$,
albeit some terms at this order still depend on the target spin in a specific manner.
\end{abstract}

\maketitle

\section{Introduction}

As is taught in virtually every quantum field theory textbook,
a generic tree-level QED process with emission of a low-energy photon exhibits simplifying feature~\cite{Schwartz:2014sze}.
In the soft limit, the full QED amplitude can be factorized into the product of the simpler one with the external
photon removed times a universal eikonal factor. The universal pattern governing the emission of soft photon can be readily carried over
to the case of emission of soft graviton~\cite{Weinberg:1965nx}, and can also be extended through the next-to-leading-order (NLO) in the small $k$ expansion
(where $k^\mu$ denotes the four-momentum of the emitted photon or graviton), which is generically referred to as the LBK theorem~\cite{Lee:1964is,Bloch:1937pw, Kinoshita:1962ur}.
Recently there has been attempt to reproduce the LBK theorem entailing soft graviton emission from the perspective of the soft-collinear effective theory~\cite{Beneke:2021umj}.

The LBK theorem only applies to the case of on-shell photon/graviton emitted from the external legs and point-like matter particles.
Nevertheless, it is also of theoretical curiosity about the soft pattern of the processes involving composite particles or the photon/graviton
emerging in the internal line, in which the LBK theorem is no longer applicable.
On the physical ground, one anticipates that in a process entailing a heavy composite particle, the soft limit implicates that
the very long wavelength of the (real or virtual) photon/graviton is unable to resolve the detailed internal structure of the composite particle,
so its proprieties can be simply summarized in terms of a few low order multipoles, correspondingly the expanded cross sections might exhibit
some simple textures.  A classical example is the soft limit of the Compton scattering,
with the leading contribution represented by the Thomson cross section, depending only on the total electric charge of the composite target,
while the NLO contribution in $1/M$ expansion becomes sensitive to its magnetic dipole~\cite{Low:1954kd,Gell-Mann:1954wra}.
Another illuminating example is the Rutherford scattering process, where a low-energy projectile bombards a static, heavy, composite target particle
bearing arbitrary spin, mediated by a $t$-channel photon exchange. Recently the soft limit of the electromagnetic Rutherford scattering process
has been comprehensively investigated by two of the authors, and some simple patterns about the target spin dependence have been
identified upon performing the heavy target mass expansion~\cite{Jia:2023udz}.

The central theme of this work is to extend the preceding analysis in electromagnetic case~\cite{Jia:2023udz} to the gravitational Rutherford scattering process,
{\it i.e.}, a structureless point-like projectile bombs on a static, heavy, composite target particle bearing arbitrary spins,
now with the graviton being the force carrier.
We consider two types of benchmark gravitational Rutherford scattering processes, {\it e.g.},
a low-energy massless structureless projectile hits a static massive composite particle with spin up to $2$,
and a non-relativistic light projectile bombs on a heavy static composite target.
The major observation of this work is similar to what is found in its electromagnetic counterpart~\cite{Jia:2023udz}:
the unpolarized cross sections in the first type are universal (independent of the target spin) at the first two orders in $1/M$ expansion,
yet differ at the next-to-next-to-leading order (NNLO), though some terms at this order still remain universal or depend on the target spin in a definite manner.
The unpolarized cross sections in the second type are universal at the lowest order in projectile velocity expansion and through all orders in $1/M$,
insensitive to both projectile and target's spin.  The universality partially breaks down at relative order-$v^2/M^2$,
though some terms at this order are still universal or depend on the target spin in a recognizable manner.

The rest of the paper is distributed as follows.
In Sec.~\ref{sec:general:amplitude}, we present the expression of the tree-level amplitude for gravitational
Rutherford scattering process involving a heavy composite spinning target particle, and specify the parametrization of the
gravitational form factor of massive target particle carrying various spin.
In Sec.~\ref{sec:soft:behavior:scenario:1}, we consider the low-energy massless point-like projectile with spin-0, ${1\over 2}$ and $1$,
striking on a heavy composite target particle with spin ranging from 0 to 2.
We organize the unpolarized cross section in the heavy target mass expansion up to NNLO,
and identify some universal pattern about the target spin dependence.
In Sec.~\ref{sec:soft:behavior:scenario:2:NR}, we consider another type of gravitational Rutherford scattering process,
where the projectile is replaced by a slowly-moving light structureless particle with spin ranging from 0 to 1.
We identify some universal pattern of the cross section in the double expansion of the projectile velocity and $1/M$.
We summarize in Sec.~\ref{sec:summary}.
In Appendix~\ref{appendix:A}, we present the spin sum formula for the target particles with various spin.
In Appendix~\ref{appendix:B}, we tackle the gravitational Rutherford scattering based on the heavy black hole effective theory (HBET),
taking the massless spinless projectile and heavy spinless target particle as example.
The reason why the NLO amplitude vanishes becomes transparent from the perspective of effective field theory.

\section{Amplitude of gravitational Rutherford scattering involving a heavy composite target particle}
\label{sec:general:amplitude}

\begin{figure}[t]
\center{
\includegraphics[scale=0.7]{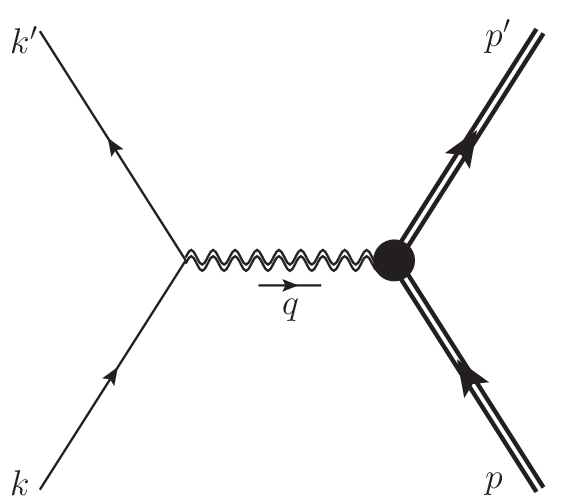}
\caption {\label{fig:feynman diag} Tree-level Feynman diagram for gravitational
Rutherford scattering process $I N\to I N$. The double wavy line represents the graviton propagator.
The thick double line represents the heavy target particle,  and the heavy dot denotes the gravitational vertex
given in \eqref{Def:GFF:composite:target:particle}.
}}
\end{figure}

In general relativity, the energy-momentum tensor of the matter field plays the role of the gravitational sources.
The symmetric Belinfante-Rosenfeld energy-momentum tensor is defined as
\beq
    T_{\mu \nu }=\frac{2}{\sqrt{-g}} \frac{\delta S}{ \delta g^{\mu \nu }},
\eeq
where $S$ denotes the Einstein-Hilbert action of the matter field, $g^{\mu\nu}$ is the spacetime metric tensor.
In the weak gravity case, one approximates $g_{\mu\nu}(x)=\eta_{\mu\nu}+\kappa h_{\mu\nu}(x)$ with $\eta_{\mu\nu}={\rm diag}(1,-1,-1,-1)$ being the Minkowski metric. Here $\kappa=\sqrt{32\pi G_N}$,
with Newton's constant $G_\text{N}=6.709\times 10^{-39}\,\mathrm{GeV}^{-2}$. In the linearized approximation,
the matter field couples with the graviton simply through the interaction $\mathcal{L}_{int}= {\kappa\over 2} h_{\mu \nu}T^{\mu \nu}$.

In this work, we focus on the gravitational Rutherford scattering process $I(k) N(p)\to I(k^\prime) N(p^\prime)$, where $I$ and $N$ represents a point-like projectile and a
heavy composite target particle, respectively. We are working in the laboratory frame where the target particle is at rest, so that the four-momentum of the target particle in the initial state
becomes $p^\mu=(M,{\bf 0})$.
As depicted in Fig.~\ref{fig:feynman diag}, the tree-level gravitational Rutherford scattering process is induced by the $t$-channel graviton exchange, and the corresponding
amplitude reads
\begin{align}
  \mathcal{M}=\frac{\kappa^{2}\mathcal{P}_{\mu\nu\rho\sigma}}{4q^2}\langle I\left(k'\right)
  \vert T^{\mu \nu} \vert I\left(k\right)\rangle\langle N\left(p',\lambda'\right)|T^{\rho \sigma} |N\left(p,\lambda\right)\rangle,
\label{Gravitational:Rutherford:scatt:amplitude}
\end{align}
with $\mathcal{P}_{\mu\nu\rho\sigma}\equiv \eta_{\mu\rho}\eta_{\nu\sigma}+\eta_{\mu\sigma}\eta_{\nu\rho}-\eta_{\mu\nu}\eta_{\rho\sigma}$ corresponding to the harmonic gauge.
$q=k-k'$ represents the momentum carried by the virtual graviton, $\lambda,\lambda'$ denote the polarization indices for the massive spinning target particle.
For simplicity, we have suppressed the spin indices of the projectile particle.

We will consider three different types of projectile particles, the spin-$0,1/2,1$ point-like particles.
With the mass denoted by $m$, the corresponding energy-momentum tensors read
\begin{subequations}
\begin{align}
T^{\mu \nu }=&\partial^{\mu }\phi \partial^{\nu}\phi-\frac{\eta^{\mu \nu}}{2}\left(\partial^{\rho}\phi\partial_{\rho}\phi-m^2\phi^2\right)  , &&\text{spin-}0
\\
T^{\mu \nu }=&\frac{i}{4}\bar{\psi}\left[\gamma^{\mu}\left(\partial^{\nu}-\overset{\leftarrow}{\partial^{\nu}} \right)+\gamma^{\nu}\left(\partial^{\mu}-\overset{\leftarrow}{\partial^{\mu}} \right)\right]\psi, &&\text{spin-}\dfrac{1}{2}
\\
T^{\mu \nu }=&-F^{\mu \lambda}F^{\nu}\,_{\lambda} +\frac{1}{4} \eta^{\mu \nu }F^{\rho \sigma}F_{\rho \sigma}-\frac{1}{2}m^2 \eta^{\mu \nu }A^{\lambda}A_{\lambda}+m^2A^{\mu}A^{\nu}. &&\text{spin-}1
\end{align}
\end{subequations}

The corresponding gravitational matrix elements involving the projectile particles can be readily worked out:
\begin{subequations}
\bqa
 \langle I(k')\vert T^{\mu \nu}\vert I(k)\rangle &=&
  k^\mu k'^\nu+k^\nu k'^\mu- \eta^{\mu\nu}(k \cdot k'-m^2),\qquad\qquad \text{spin-}0
\label{GFF:point:like:projectile:spin:0}
\\
\langle I(k')\vert T^{\mu \nu} \vert I(k)\rangle &=& \frac{1}{4}\bar{u}(k')\left(\gamma^{\mu}(k'^{\nu}+k^{\nu})+\gamma^{\nu}(k'^{\mu}+k^{\mu})\right)u(k),
\qquad\qquad \text{spin-}\dfrac{1}{2}
    \\
\langle I(k')\vert T^{\mu \nu}\vert I(k)\rangle &=&  \varepsilon^*_{\sigma}(k')[P^{\mu\nu\rho\sigma}(k\cdot k'-m^2) +
\eta^{\mu\nu}k^\sigma  k'^{\rho}+ \eta^{\rho\sigma}\left(k^\mu k'^{\nu}+k^\nu k'^{\mu}\right) \qquad\qquad \text{spin-}1
\nn\\
&-& \eta^{\nu\rho}k^\sigma k'^{\mu}-\eta^{\mu\rho}k^\sigma k'^{\nu}-\eta^{\nu\sigma}k^\mu k'^{\rho}-\eta^{\mu\sigma}k^\nu k'^{\rho}]\varepsilon_{\rho}(k).
\eqa
\end{subequations}

The gravitational matrix elements involving the composite target particles in \eqref{Gravitational:Rutherford:scatt:amplitude} are
in general nonperturbative objects, which vary with target species. In literature they are usually referred to as the {\it gravitational form factors} (GFFs)~\cite{Kobzarev:1962wt, Pagels:1966zza}.
Since the GFFs encode some essential mechanical properties of a hadron such as mass, spin and shear force distributions\cite{Polyakov:2002yz, Polyakov:2018exb, Polyakov:2018zvc}, people's interest toward hadron's GFFs has revived in recent years. Although it is unfeasible to detect the gravitational Rutherford scattering in the foreseeable future, it is of high priority of the current and forthcoming $ep$ facilities
such as {\tt Jlab}, {\tt EIC} and {\tt EicC} to  extract nucleon' GFFs in an indirect way~\cite{Burkert:2023wzr}~\footnote{Note that it has been recently proposed that the nucleon GFF may be accessed
in the future electron-ion collider via the interference effect between the photon-induced and the massive-graviton-induced amplitude in some beyond Standard Model scenarios~\cite{Hatta:2023fqc}.}.
Recently the GFFs of the proton and pion at small momentum transfer have been investigated from the lattice QCD simulation~\cite{Hackett:2023nkr, Hackett:2023rif}.

In this work we consider five types a composite target particles of mass $M$, with spin varying from 0 to 2.
In line with the Lorentz group representation, the corresponding gravitational matrix elements involving various target particles
can be decomposed into the linear combination of different GFFs~\cite{Cotogno:2019vjb}~\footnote{Current conservation enables one to drop the $F_{30}$ and $F_{6,i}$ terms,
as well as implies $F_{31}+F_{50}=0$~\cite{Cotogno:2019vjb}.}:
\begin{subequations}
\begin{align}
\langle N\left(p',\lambda'\right)|T^{\mu \nu}|N\left(p,\lambda\right)\rangle_{s=0}&=
2P^\mu P^\nu\,F_{10}\left(\dfrac{q^2}{M^2}\right)
+2\left(q^\mu q^\nu-\eta^{\mu\nu}q^2\right)F_{20}\left(\dfrac{q^2}{M^2}\right),
\\
\langle N\left(p',\lambda'\right)|T^{\mu \nu}|N\left(p,\lambda\right)\rangle_{s=\frac{1}{2}}&= \overline u(p',\lambda')\Big[
 2P^\mu P^\nu\,F_{10}\left(\dfrac{q^2}{M^2}\right)
+2\left( q^\mu q^\nu-\eta^{\mu\nu} q^2\right)F_{20}\left(\dfrac{q^2}{M^2}\right)
\nn\\
& +P^{\{\mu}\tfrac{i}{2}\sigma^{\nu\}\rho} q_\rho F_{40}\left(\dfrac{q^2}{M^2}\right)
\Big]u(p,\lambda),
\\
\langle N\left(p',\lambda'\right)|T^{\mu \nu}|N\left(p,\lambda\right)\rangle_{s=1}=&-\varepsilon^*_{\alpha'}(p',\lambda')\bigg[
2P^ \mu P^\nu \left(\eta^{\alpha'\alpha}\,F_{10}\left(\dfrac{q^2}{M^2}\right)-\frac{ q^{\alpha'} q^\alpha}{2M^2}\,F_{11}\left(\dfrac{q^2}{M^2}\right)\right)
\nn\\
& +2\left( q^\mu q^\nu-\eta^{\mu\nu} q^2\right)\left(\eta^{\alpha'\alpha}\,F_{20}\left(\dfrac{q^2}{M^2}\right)-\frac{ q^{\alpha'} q^\alpha}{2M^2}\,F_{21}\left(\dfrac{q^2}{M^2}\right)\right)
\nn\\
& -2M^2\eta^{\mu\nu}\left(\frac{ q^{\alpha'} q^\alpha}{2M^2}\,F_{31}\left(\dfrac{q^2}{M^2}\right) \right)-P^{\{\mu}\eta^{\nu\}[\alpha'} q^{\alpha]}F_{40}\left(\dfrac{q^2}{M^2}\right)
\nn\\
& -\left( q^{\{\mu}\eta^{\nu\}\{\alpha'} q^{\alpha\}}-\eta^{\mu\nu} q^{\alpha'} q^{\alpha}-\eta^{\alpha'\{\mu}\eta^{\nu\}\alpha} q^2\right)F_{50}\left(\dfrac{q^2}{M^2}\right)\bigg]\varepsilon_{\alpha}(p,\lambda),
\\
\langle N\left(p',\lambda'\right)\vert T^{\mu \nu}|N\left(p,\lambda\right)\rangle_{s=\frac{3}{2}}=&-\overline u_{\alpha'}(p',\lambda')\bigg[
2P^ \mu P^\nu \left(\eta^{\alpha'\alpha}\,F_{10}\left(\dfrac{q^2}{M^2}\right)-\frac{ q^{\alpha'} q^\alpha}{2M^2}\,F_{11}\left(\dfrac{q^2}{M^2}\right)\right)
\nn\\
&+2\left( q^\mu q^\nu-\eta^{\mu\nu} q^2\right)\left(\eta^{\alpha'\alpha}\,F_{20}\left(\dfrac{q^2}{M^2}\right)-\frac{ q^{\alpha'} q^\alpha}{2M^2}\,F_{21}\left(\dfrac{q^2}{M^2}\right)\right)
\nn\\
&-2M^2\eta^{\mu\nu}\left(\frac{ q^{\alpha'} q^\alpha}{2M^2}\,F_{31}\left(\dfrac{q^2}{M^2}\right) \right)\nonumber\\
&+P^{\{\mu}\tfrac{i}{2}\sigma^{\nu\}\rho} q_\rho \left(\eta^{\alpha'\alpha}\,F_{40}\left(\dfrac{q^2}{M^2}\right)-\frac{ q^{\alpha'} q^\alpha}{2M^2}\,F_{41}\left(\dfrac{q^2}{M^2}\right)\right)
\nn\\
&-\left( q^{\{\mu}\eta^{\nu\}\{\alpha'} q^{\alpha\}}-\eta^{\mu\nu} q^{\alpha'} q^{\alpha}-\eta^{\alpha'\{\mu}\eta^{\nu\}\alpha} q^2\right)F_{50}\left(\dfrac{q^2}{M^2}\right)\bigg]u_{\alpha}(p,\lambda),
\\
\langle N\left(p',\lambda'\right)| T^{\mu \nu} |N\left(p,\lambda\right)\rangle_{s=2}=&\varepsilon^*_{\alpha'_1\alpha'_2}(p',\lambda')\bigg[
2P^\mu P^\nu\left(\eta^{\alpha'_1\alpha_1}\eta^{\alpha'_2\alpha_2}\,F_{10}\left(\dfrac{q^2}{M^2}\right)-\frac{ q^{\alpha'_1} q^{\alpha_1}}{2M^2}\,\eta^{\alpha'_2\alpha_2}\,F_{11}\left(\dfrac{q^2}{M^2}\right)\right.
\nn\\
&\left.+\frac{ q^{\alpha'_1} q^{\alpha_1}}{2M^2}\,\frac{ q^{\alpha'_2} q^{\alpha_2}}{2M^2}\,F_{12}\left(\dfrac{q^2}{M^2}\right)\right)
\nn\\
&+2\left( q^\mu q^\nu-\eta^{\mu\nu} q^2\right)\left(\eta^{\alpha'_1\alpha_1}\eta^{\alpha'_2\alpha_2}\,F_{20}\left(\dfrac{q^2}{M^2}\right)-\frac{ q^{\alpha'_1} q^{\alpha_1}}{2M^2}\,\eta^{\alpha'_2\alpha_2}\,F_{21}\left(\dfrac{q^2}{M^2}\right)\right.
\nn\\
&\left.+\frac{ q^{\alpha'_1} q^{\alpha_1}}{2M^2}\,\frac{ q^{\alpha'_2} q^{\alpha_2}}{2M^2}\,F_{22}\left(\dfrac{q^2}{M^2}\right)\right)
\nn\\
&-2M^2\eta^{\mu\nu}\left(\frac{ q^{\alpha'_1} q^{\alpha_1}}{2M^2}\,\eta^{\alpha'_2\alpha_2}\,F_{31}\left(\dfrac{q^2}{M^2}\right) -
\frac{ q^{\alpha'_1} q^{\alpha_1}}{2M^2}\,\frac{ q^{\alpha'_2} q^{\alpha_2}}{2M^2}\,F_{32}\left(\dfrac{q^2}{M^2}\right)\right)
\nn\\
&-P^{\{\mu}\eta^{\nu\}[\alpha'_2} q^{\alpha_2]}\left(\eta^{\alpha'_1\alpha_1}F_{40}\left(\dfrac{q^2}{M^2}\right)-\frac{ q^{\alpha'_1} q^{\alpha_1}}{2M^2}F_{41}\left(\dfrac{q^2}{M^2}\right)\right)
\nn\\
&-\left( q^{\{\mu}\eta^{\nu\}\{\alpha'_2} q^{\alpha_2\}}-\eta^{\mu\nu} q^{\alpha'_2} q^{\alpha_2}-\eta^{\alpha'_2\{\mu}\eta^{\nu\}\alpha_2} q^2\right)
\nn\\
&\times \left(\eta^{\alpha'_1\alpha_1}F_{50}\left(\dfrac{q^2}{M^2}\right)-\frac{ q^{\alpha'_1} q^{\alpha_1}}{2M^2}F_{51}\left(\dfrac{q^2}{M^2}\right)\right)
\nn\\
&+ q^{[\alpha'_2}\eta^{\alpha_2]\{\mu}\eta^{\nu\}[\alpha'_{1}} q^{\alpha_{1}]}F_{70}\left(\dfrac{q^2}{M^2}\right)\bigg]\varepsilon_{\alpha_1\alpha_2}(p,\lambda),
\end{align}
\label{Def:GFF:composite:target:particle}
\end{subequations}
where $P\equiv (p+p')/2$ is the average momentum of the target particle between the initial and final states,
$q \equiv p'-p$ denotes the transfer momentum. $a^{\{ \mu} b^{\nu \} }\equiv a^{\mu}b^{\nu}+a^{\nu}b^{\mu}$, $a^{[  \mu} b^{\nu ] } \equiv a^{\mu}b^{\nu}-a^{\nu}b^{\mu}$, and $\sigma^{\mu \nu}={i\over 2}[\gamma^{\mu},\gamma^{\nu}]$.
$u$, $\varepsilon^\mu$, $u^\mu$, $\varepsilon^{\alpha\beta}$ signify the wave functions of the spin-${1/2}$, 1, ${3/2}$, and 2 particles, respectively.
Note various GFFs are normalized to be dimensionless Lorentz scalars that solely depend on the ratio $q^2/M^2$.
Note that we have suppressed terms that are forbidden by the current conservation.

From \eqref{Def:GFF:composite:target:particle} one observes that for
target particle with spin $s$, the number of independent GFFs is $2(s+1)+3\lfloor s\rfloor-\Theta(s-1)$~\cite{Cotogno:2019vjb}~\footnote{
The Heaviside step function is defined as $\Theta(x)=1$ for $x\geq 0$, otherwise vanishes. The symbol
$\lfloor s \rfloor$ signifies the floor function of $s$. Later the symbol $\lceil s\rceil$ will be used to represent the
ceiling function of $s$.}. Analogous to electromagnetic form factors, various GFFs with zero momentum transfer encapsulate the
properties of the gravitational multipoles of the composite target particles.
For the target particle carrying an arbitrary spin $s$,  the $F_{10}$ and $F_{40}$ have an absolute normalization at zero momentum transfer,
$F_{10}(0)=1$ and $F_{40}(0)=s$,  as dictated by the energy-momentum and angular momentum conservation, respectively.
The $F'_{10}(0)$~\footnote{The Taylor expansion of the GFF around the origin is understood to be $F_n (q^2/M^2)= F_n(0)+ F_n'(0){q^2/M^2}+
{\mathcal O}(1/M^4)$.}, $F_{11}(0)$, $F_{20}(0)$ and $F_{50}(0)$ terms are related to angular momentum~\cite{Leader:2013jra, Polyakov:2002yz}, pressure and shear force~\cite{Polyakov:2002yz, Polyakov:2018exb, Polyakov:2018zvc} of the target particle. The mass radius of the target particle can also be obtained from the linear combination of these form factors with zero momentum transfer.

\section{Low-energy Gravitational Rutherford scattering with massless projectile }
\label{sec:soft:behavior:scenario:1}

We first consider the case of the massless projectile of spin $j$.
The corresponding differential unpolarized cross section in the laboratory frame is given by
\begin{align}
    &\dfrac{\mathrm{d}\sigma}{\mathrm{d}\cos\theta}=
    \dfrac{1}{2 |{\bf k}|}\cdot\dfrac{1}{2M}\cdot\dfrac{{\bf k'}^2}{8\pi|{\bf k}| M}\cdot {1\over 2j+1}\dfrac{1}{2s+1}\sum_\text{spins}\left|\mathcal{M}\right|^2,
\label{def:cross:section:Rutherford:massless:projectile}
\end{align}
where $\theta$ denotes the polar angle between the reflected  and incident projectile.
The magnitude of the three-momentum of the outgoing projectile, $|{{\bf k}^\prime}|$,
can be expressed in terms of $|{\bf k}|$, $M$, and $\cos\theta$:
\beq
|{\bf k}'|= {|{\bf k}| \over 1+{|{\bf k}|\over M}\left ( 1-\cos\theta\right)}.
\label{kprime:k:relation}
\eeq

Squaring the amplitude in \eqref{Gravitational:Rutherford:scatt:amplitude},
summing over the polarization in the final state and averaging upon the polarizations
in the initial state utilizing the spin sum formula in Appendix for target particles,
one encounters rather lengthy and cumbersome-looking expressions.
It is difficult to identify any clear pattern about the dependence on the heavy target particle spin.
Nevertheless, since we are solely concerned with the low-energy limit $|{\bf k}| \ll M$,
it becomes elucidating to carry out the heavy target mass expansion for the unpolarized cross sections. As we will see,
the soft behavior of the gravitational Rutherford scattering becomes transparent and one
is able to recognize some universal patterns.

\subsection{Massless spin-$0$ projectile}
\label{massless:spin:0:projectile}

Expanding \eqref{def:cross:section:Rutherford:massless:projectile} in powers of $1/M$, one immediately observes that the first two terms of
the unpolarized cross sections are universal, {\it e.g.}, independent of the target particle spin:
\begin{subequations}
\bqa
 &&\left(\dfrac{\mathrm{d}\sigma}{\mathrm{d}\cos\theta}\right)_\text{LO}^{s}= {\kappa^{4} M^2 F_{10}^2  \over 512 \pi \sin^4 \frac{\theta}{2}},
 \label{cross:section:Rutherford:massless:spin:0:projectile:LO}
 \\
 &&\left(\dfrac{\mathrm{d}\sigma}{\mathrm{d}\cos\theta}\right)_\text{NLO}^{s}=  -{ \kappa^4 M |{\bf k}| F_{10}^2\over 128\pi \sin^2\frac{\theta}{2}},\label{cross:section:Rutherford:massless:spin:0:projectile:NLO}
\eqa
\label{cross:section:Rutherford:massless:spin:0:projectile:LO:NLO}
\end{subequations}
with the occurring GFFs evaluated at the zero momentum transfer.
For simplicity, we have adopted $F_n$ as the shorthand for $F_n(0)$ from now on.
Note that $F_{10}=1$ for any type of composite target particles.
The leading order (LO) term is identical to the cross section obtained from the light-bending angle in classical general relativity\cite{Donoghue:1986ya,Bjerrum-Bohr:2014zsa}.
This is intuitively as expected,  because in the soft limit, the long-wavelength graviton can only feel the total mass of the composite target particle,
insensitive to any further details about its internal structure.
Interestingly, the next-to-leading order (NLO) term still remains universal, which originates from expanding the phase factor
factor ${{\bf k}'}^2/{\bf k}^2$ to NLO in $1/M$.
It is instructive to understand why only a single GFF $F_{10}$ contributes at NLO
from the angle of effective field theory.
We will devote Appendix~\ref{appendix:B} to such an analysis.

In contrast, at the next-to-next-to-leading-order (NNLO) in heavy target mass expansion, the differential cross sections starts to depend on the target particle spin:
\begin{align}
   \left(\dfrac{\mathrm{d}\sigma}{\mathrm{d}\cos\theta}\right)_\text{NNLO}^{s}
    =&-\frac{\kappa^4 {\bf k}^2}{64\pi \sin^2 \frac{\theta}{2}}
    \left\{{F_{10} F'_{10}}-\dfrac{1}{2}F_{10}F_{20}\left(1-\cos\theta\right)
    +\dfrac{1}{8}F_{10}^2\left[7\cos{\theta}-\dfrac{2}{3}\left(\dfrac{21}{2}+s+\lceil s\rceil\right)\right]
    \right.
\nn\\
 &+\Theta\left(s-\frac{1}{2}\right)\left[\dfrac{(-1)^{2s}+7}{24}F_{10}F_{40}
    -f^{(0)}_s \,F_{40}^2(\cos{\theta}+1)\right]
\nn\\
    & -\dfrac{1}{6}\Theta\left(s-1\right)
    \left[2\cos{\theta}F_{10}F_{50}+F_{10}F_{11} \right]
\nn\\
 & -\left.\Theta\left(s-2\right)\dfrac{1}{6}F_{10}F_{70}(1+\cos{\theta})\right\}, \qquad s=0,\dfrac{1}{2},1,\dfrac{3}{2},2
 \label{eq: cross section-NNLO}
\end{align}
with
\begin{subequations}
\begin{align}
f^{(0)}_{\frac{1}{2}} =&\dfrac{1}{16},
\\
f^{(0)}_{1} =&\dfrac{1}{6},
 \\
f^{(0)}_{\frac{3}{2}}  =&\dfrac{5}{144},
\\
f^{(0)}_{2}= &\dfrac{1}{8}.
\end{align}
\end{subequations}

We observe that $F'_{10}F_{10}$, $F_{10} F_{20}$, $F_{10}^2\cos{\theta}$ terms still remain universal, {\it i.e.},
independent of the target spin.
In fact, the $F'_{10}F_{10}$ terms
actually have the same origin of the LO and NLO cross sections, which correspond to higher-order term
in Taylor expansion of $F^2_{10}(q^2/M^2)$ in the squared LO amplitude and the phase space measure.
The GFFs $F'_{10}$, $F_{11}$, $F_{20}$ and $F_{50}$ reflect the mechanical properties of the composite target particles such as angular momentum~\cite{Leader:2013jra, Polyakov:2002yz}, pressure and shear force~\cite{Polyakov:2002yz, Polyakov:2018exb, Polyakov:2018zvc}.
Our results indicate that at NNLO in heavy target mass expansion,
the cross section starts to depend on the detailed three-dimensional internal structure of target particle other than its mass.
Curiously, the coefficient of the $F_{10}F_{40}$ term seems to reflect the spin-statistic characteristic of the target particle,
which alternates from $1$ (fermions) to $4/3$ (bosons).
Although we only enumerate five different kinds of target spin,
it is conceivable that the aforementioned patterns should hold true for arbitrary target spin.

\subsection{Massless spin-$1/2$ projectile}
\label{massless:spin:one:half:projectile}

We can repeat our investigation in Sec.~\ref{massless:spin:0:projectile} by replacing the projectile with a point-like massless Dirac fermion.
After conducting the heavy target mass expansion, we again observe that the unpolarized cross sections exhibit universal forms at LO and NLO in $1/M$ expansion:
\begin{subequations}
\bqa
 &&\left(\dfrac{\mathrm{d}\sigma}{\mathrm{d}\cos\theta}\right)_\text{LO}^{s}= {\kappa^{4} M^2  F_{10}^2 \cos^2\frac{\theta}{2} \over 512 \pi \sin^4 \frac{\theta}{2}},
 \\
 &&\left(\dfrac{\mathrm{d}\sigma}{\mathrm{d}\cos\theta}\right)_\text{NLO}^{s}=  -{\kappa^4 M |{\bf k}| F_{10}^2  \cos^2\frac{\theta}{2}  \over 128\pi \sin^2 \frac{\theta}{2}}.
\eqa
\label{cross:section:Rutherford:massless:spin:one:half:projectile:LO:NLO}
\end{subequations}
It is worth emphasizing that, although these expressions are independent of the target spin,
they are sensitive to the projectile spin.
To be definite, there arises an extra factor $\cos^2 \frac{\theta}{2}$ in the numerators of \eqref{cross:section:Rutherford:massless:spin:one:half:projectile:LO:NLO}
with respect to \eqref{cross:section:Rutherford:massless:spin:0:projectile:LO:NLO} in the case of spinless projectile.

As in Sec.~\ref{massless:spin:0:projectile}, the universality becomes partially violated at NNLO.
For various composite target particles with spin ranging from $0$ to $2$, the NNLO contributions to the cross sections become
\begin{align}
\left(\dfrac{\mathrm{d}\sigma}{\mathrm{d}\cos\theta}\right)_\text{NNLO}^{s}
=&\frac{\kappa^4 {\bf k}^2}{64\pi \sin^2 \frac{\theta}{2} }
    \left\{{-F_{10} F'_{10}}\cos^2 \frac{\theta}{2}
    +\dfrac{1}{8}F_{10}^2\cos^2\frac{\theta}{2}\left[13\sin^2\frac{\theta}{2}+\dfrac{2}{3}\left(s+\lceil s\rceil\right)\right]
    \right.
\nn\\
    &-\Theta\left(s-\frac{1}{2}\right)\left[\dfrac{(-1)^{2s}+7}{24}F_{10}F_{40}\cos^2\frac{\theta}{2}
    - f^{({1\over 2})}_{s} \, F_{40}^2\left(\cos{2\theta}+8\cos{\theta}+23\right)\right]
\nn\\
    &\left.+\dfrac{1}{6}\Theta\left(s-1\right)
    F_{10}\left(2F_{50}+F_{11}\right)\cos^2\frac{\theta}{2}
    +\Theta\left(s-2\right)\dfrac{1}{3}F_{10}F_{70}\cos^2 \frac{\theta}{2}\right\}  ,\qquad s=0,\dfrac{1}{2},1,\dfrac{3}{2},2
\end{align}
with
\begin{subequations}
  \begin{align}
   f^{({1\over 2})}_{\frac{1}{2}} =& {1\over 256},
\label{eq: cross section-NNLO-1/2}
\\
  f^{({1\over 2})}_{1}=& {1\over 96},
\\
f^{({1\over 2})}_{\frac{3}{2}} = & {5\over 2304},
\\
f^{({1\over 2})}_{2} =  & {1\over 128}.
\end{align}
\end{subequations}
Similar to the pattern revealed in the case of massless spinless projectile, we observe that the coefficients of the $F'_{10}F_{10}$, $F_{10} F_{20}$,
$F_{10}^2\sin^2{\frac{\theta}{2}}\cos^2{\frac{\theta}{2}}$ remain independent of the target spin.
The $F'_{10}F_{10}$ and $F_{10}^2\cos\theta$ terms actually have the same origin of the LO and NLO cross sections.
The coefficient of the $F_{10}F_{40}$ term again reflects the spin-statistic characteristic of the target particle,
which alternates from $1$ (fermions) to $4/3$ (bosons).

\subsection{Massless spin-$1$ projectile}
\label{massless:spin:1:projectile}

We can repeat the the preceding analysis by turning to a massless spin-$1$ projectile.
Upon heavy target mass expansion, analogous to the cases of the spin-0 and spin-${1\over 2}$ projectile,
we again observe the similar universal pattern at LO and NLO:
\begin{subequations}
\bqa
 &&\left(\dfrac{\mathrm{d}\sigma}{\mathrm{d}\cos\theta}\right)_\text{LO}^{s}= {\kappa^{4} M^2  F_{10}^2 \cos^4\frac{\theta}{2} \over 512 \pi \sin^4 \frac{\theta}{2}},
 \\
 &&\left(\dfrac{\mathrm{d}\sigma}{\mathrm{d}\cos\theta}\right)_\text{NLO}^{s}=  -{\kappa^4 M |{\bf k}| F_{10}^2  \cos^4\frac{\theta}{2}  \over 128\pi \sin^2 \frac{\theta}{2}},
\eqa
\label{cross:section:Rutherford:massless:spin:1:projectile:LO:NLO}
\end{subequations}
except there appears an extra factor $\cos^4 \frac{\theta}{2}$ in the numerators with respect to the case of spinless projectile.

The universality becomes partially violated at NNLO.
For target particle with spin ranging from 0 to 2, the NNLO contributions to the unpolarized cross sections become
\begin{align}
  \left(\dfrac{\mathrm{d}\sigma}{\mathrm{d}\cos\theta}\right)_\text{NNLO}^{s}=&
    -\frac{\kappa^4 {\bf k}^2\cos^2\frac{\theta}{2}}{256\pi\sin^2\frac{\theta}{2}} \left\{F_{10}^2\cos^2\frac{\theta}{2}
    \left[3\cos\theta-\dfrac{1}{3}\left(9-s-\lceil s\rceil\right)\right] +4F_{10}F_{10}'\cos^2\frac{\theta}{2}\right.
\nn\\
& +\Theta\left(s-\frac{1}{2}\right)\left[\dfrac{(-1)^{2s}+7}{6} F_{10}F_{40}\cos^2\dfrac{\theta}{2}+ f^{(1)}_s \, F_{40}^2(\cos{\theta}-3)\right]
\nn\\
    &\left.-\dfrac{2}{3}F_{10}\left[\Theta\left(s-1\right)\left(F_{11}+2F_{50}\right)+\Theta\left(s-2\right)2F_{70}\right]\cos^2\dfrac{\theta}{2}\right\},\qquad s=0,\dfrac{1}{2},1,\dfrac{3}{2},2
\end{align}
with
\begin{subequations}
\begin{align}
f^{(1)}_{\frac{1}{2}}  =&\dfrac{1}{4},
\\
f^{(1)}_{1}=&\dfrac{2}{3},
\\
f^{(1)}_{\frac{3}{2}} =&\dfrac{5}{36},
\\
f^{(1)}_{2}=& \dfrac{1}{2}.
\end{align}
\end{subequations}
Similar to the projectile of spin-0 and ${1\over 2}$, we observe that the coefficients of
$F'_{10}F_{10}$, $F_{10} F_{20}$, $F_{10}^2\cos{\theta}\cos^2{\frac{\theta}{2}}$ terms are independent of the target spin.
Actually the $F'_{10}F_{10}$ and $F_{10}^2\cos^3\theta$ terms have the same origin of the LO and NLO cross sections.
Again, the coefficient of the $F_{10}F_{40}$ term again reflects the spin-statistic characteristic of the target particle,
which alternates from $1$ (fermions) to $4/3$ (bosons).

\section{Gravitational Rutherford scattering with non-relativistic light projectile }
\label{sec:soft:behavior:scenario:2:NR}

In this section, we turn to the original prototype of Rutherford scattering process, {\it i.e.}, a slowly moving light projectile hits a heavy static target,
albeit with the interaction mediated by exchanging a graviton exchange rather than exchanging a photon.
We assume the projectile is point-like particle with mass $m\ll M$.

The differential cross section for the nonrelativistic Rutherford scattering in the laboratory frame is given by
\begin{align}
\dfrac{\mathrm{d}\sigma}{\mathrm{d}\cos\theta}=\dfrac{1}{32\pi M}\left[{p'^0+k'^0\left(1-\dfrac{|\mathbf{k}|}{|\mathbf{k}'|}\cos\theta\right)}\right]^{-1}\dfrac{|\mathbf{k}'|}
{|\mathbf{k}|} {1\over 2j+1}\dfrac{1}{2s+1}\sum_\text{spins}\left|\mathcal{M}\right|^2,
\label{def:diff:cross:section:massless:lab:frame}
\end{align}
where the projectile bears spin $j$, $\theta$ denotes the polar angle between the reflected and incident projectile.

Similar to Sec.~\ref{sec:soft:behavior:scenario:1}, let us consider again three different types of projectiles and five different types of targets.
The resulting expressions of unpolarized cross sections are generally rather lengthy and unilluminating.
Since there are three widely separated scales inherent in this process, ${\bf k}\ll m \ll M$,
the appropriate way of extracting the soft behavior is to simultaneously expand the differential cross sections in powers of
$v=|{\bf k}|/m$ (velocity of the projectile) and $m/M$.
The necessity of conducting double expansion renders this case more complicated than the case of
low-energy massless projectile as discussed in Sec.~\ref{sec:soft:behavior:scenario:1}.

Interestingly, at the lowest order in velocity yet to all orders in $1/M$,
the differential cross sections scales assumes a uniform form:
\begin{align}
\left(\dfrac{\mathrm{d}\sigma}{\mathrm{d}\cos\theta}\right)^{s}_{(v^0)}=& \dfrac{ F_{10}^2 \kappa^4  m^4 M(M+m)^2 \left(\sqrt{M^2-m^2\sin^2\theta}+m \cos\theta\right)^2}{512\pi \mathbf{k}^4\sqrt{M^2-m^2\sin^2\theta}\left(M-\cos\theta \sqrt{M^2-m^2\sin^2\theta}+m \sin^2\theta\right)^2}
\nn\\
= & \frac{\kappa^4 M^2 m^4 F_{10}^2}{2048\pi{\bf k}^4\sin^4\frac{\theta}{2}}+
\mathcal{O}\left({m^6\over M^4 {\bf k}^4}\right).
\label{eq: NR-LO}
\end{align}
which scales as $1/{\bf k}^4$, exactly identical to the familiar Rutherford formula obtained from the classical mechanics with Newtonian's gravitational law~\cite{Misner:1973prb}.
Note this expression is insensitive to both target and projectile's spins, since the spin degree of freedom decouples in the nonrelativistic limit. This is in contrast to the LO
expressions in the case of massless projectile in heavy target mass expansion, which is insensitive to the target spin yet depends on the projectile spin.

At NLO in velocity expansion, the differential cross sections scale as $1/{\bf k}^2$, whose explicit expressions are still rather complicated yet
vary with different projectile and target species. Nevertheless, once the heavy target mass expansion is
performed, some clear pattern starts to emerges.
In the following, we will consider three different types of light projectiles with spin ranging from 0 to 1.

\subsection{Slowly-moving spin-$0$ projectile}
\label{NR:spin:0:projectile}

At relative order-$v^2$, after conducting the heavy target mass expansion, the differential unpolarized cross section becomes particularly simple:
\begin{align}
  \left(\dfrac{\mathrm{d}\sigma}{\mathrm{d}\cos\theta}\right)^s_{(v^2)}= &
  {\kappa^4 M^2 m^2 F_{10} \over 256\pi \mathbf{k}^2 \sin^2\frac{\theta}{2}}\left[ \frac{F_{10}}{2\sin^2\frac{\theta}{2}}- {m\over M} F_{10}
  + {m^2\over M^2} g^{(0)}_s +\cdots \right],
\end{align}
with
\begin{subequations}
  \begin{align}
 g^{(0)}_s =&-F_{20}-F'_{10}
  + \dfrac{F_{10}}{4}\left[\dfrac{1}{3}\left(3+s+\lceil s\rceil\right)-2\cos\theta\right]
  -\Theta\left(s-\frac{1}{2}\right)\dfrac{1}{24}\left[(-1)^{2s}+7\right]F_{40}
\nn\\
  &+\Theta\left(s-1\right)\left(\frac{1}{6}F_{11}+F_{50}\right)
  +\Theta\left(s-2\right)\dfrac{2}{3}F_{70}, \qquad s=0,\dfrac{1}{2},1,\dfrac{3}{2},2.
\end{align}
\end{subequations}
Clearly the ${\cal O}(v^2/M^n)$ ($n=0,1$) terms remain independent of the target spin. At ${\cal O}(v^2/M^2)$, the universality becomes partially violated. Nevertheless,
the $F'_{10}$, $F_{20}$, and $F_{10}\cos\theta$ terms still do not depend on the target particle spin.
The coefficient of the $F_{40}$ term again reflects the spin-statistic characteristic of the target particle,
which alternates from $1$ (fermions) to $4/3$ (bosons).

\subsection{Slowly-moving spin-${1\over 2}$ projectile}
\label{NR:spin:one:half:projectile}

We can repeat our investigation by replacing the projectile with a slowly-moving Dirac fermion.
The ${\cal O}(v^0)$ cross section is given by \eqref{eq: NR-LO}, since the spin degree of freedom decouples in the nonrelativistic limit.
At relative order-$v^2$, after carrying out the heavy target mass expansion, the differential cross section again possesses a simple form:
\begin{align}
\left(\dfrac{\mathrm{d}\sigma}{\mathrm{d}\cos\theta}\right)^{s}_{(v^2)}
= &\dfrac{\kappa^4 M^2 m^2 F_{10}}{256\pi {\bf k}^2 \sin^2\frac{\theta}{2}}\left[\frac{F_{10}}{16}
\frac{3\cos{\theta}+5}{\sin^2\frac{\theta}{2}}-\frac{m F_{10}}{8 M} (3\cos{\theta}+5)-{ m^2\over M^2 } g^{({1\over 2})}_s +\cdots \right],
\end{align}
with
\begin{subequations}
\begin{align}
 g^{({1\over 2})}_s
    =&
    { F'_{10}}+F_{20}
    +\dfrac{1}{48}F_{10}\left[15\cos{\theta}-4\left(\dfrac{3}{4}+s+\lceil s\rceil\right)\right]
   +\Theta\left(s-\frac{1}{2}\right)\dfrac{1}{24}\left[(-1)^{2s}+7\right]F_{40}
\nn\\
    &-\Theta\left(s-1\right)
    \left(F_{50}+\dfrac{1}{6}F_{11} \right)
    -\Theta\left(s-2\right)\dfrac{2}{3}F_{70},\qquad s=0,\dfrac{1}{2},1,\dfrac{3}{2},2.
\end{align}
\label{eq:NR-NNLO}
\end{subequations}
The ${\cal O}(v^2/M^n)$ ($n=0,1$) terms remain universal.
The universality has been partially violated in the ${\cal O}(v^2/M^2)$ term.
However, even at this order, the $F'_{10}$, $F_{20}$, and $F_{10}\cos\theta$ terms still appears to be independent of the
target particle spin. The coefficient of the $F_{40}$ term again reflects the spin-statistic characteristic of the target particle,
which alternates from $1$ (fermions) to $4/3$ (bosons).

\subsection{Slowly-moving spin-$1$ projectile}
\label{NR:spin:1:projectile}

As a final example, we consider a slowly-moving light spin-$1$ projectile.
Needless to say, the ${\cal O}(v^0)$ differential cross section is again described by \eqref{eq: NR-LO}.
At relative order-$v^2$, after carrying out the heavy target mass expansion,
the expanded differential cross sections bear the following structure:
\begin{align}
\left(\dfrac{\mathrm{d}\sigma}{\mathrm{d}\cos\theta}\right)^s_{(v^2)}=
  \frac{\kappa^4  M^2 m^2 F_{10}}{1536\pi \mathbf{k}^2  \sin^2\frac{\theta}{2}}
  \left[ \frac{F_{10}\left(2\cos\theta+1\right)}{\sin^2\frac{\theta}{2}}-{2m\over M} \left(2\cos\theta+1\right) F_{10}
  -{m^2\over M^2} g^{(1)}_s +\cdots \right],
\end{align}
where
\begin{subequations}
  \begin{align}
g^{(1)}_s =&F_{10}\left[\cos\theta+\dfrac{1}{2}\left(1-s-\lceil s\rceil\right)\right]+6F_{10}'+6F_{20}+\Theta\left(s-\frac{1}{2}\right)\dfrac{1}{4}\left[(-1)^{2s}+7\right]F_{40}
\nn\\
&-\Theta\left(s-1\right)\left(F_{11}-6F_{50}\right)-\Theta(s-2)4F_{70},
\qquad s=0,\dfrac{1}{2},1,\dfrac{3}{2},2.
\end{align}
\end{subequations}
Clearly the ${\cal O}(v^2/M^n)$ ($n=0,1$) terms remain universal.
At ${\cal O}(v^2/M^2)$, the universality becomes partially violated, notwithstanding
that the $F'_{10}$, $F_{20}$, and $F_{10}\cos\theta$ terms still do not depend on the target spin.
The coefficient of the $F_{40}$ term again reflects the spin-statistic characteristic of the target particle,
which alternates from $1$ (fermions) to $4/3$ (bosons).

\section{Summary}\label{sec:summary}

In this work, we have conducted a comprehensive study of the soft pattern of the tree-level
gravitational Rutherford scattering processes.
Two classes of Rutherford scattering processes have been considered, {\it i.e.},
a low-energy massless projectile strikes on a static, heavy, composite target carrying spin up to 2,
and a slowly-moving light structureless projectile bombs on a static, heavy, spinning composite target particle.

The soft limits of both classes of gravitational Rutherford scattering processes have exhibited some universal and simple patterns.
For the former type, given a massless projectile with a certain spin, the first two terms in the heavy target mass expansion remain universal,
and the NNLO term starts to develop target spin dependence.
Nevertheless, several terms at NNLO still remain universal or have some definite pattern of dependence on the target spin.
For the latter, one has to carry out the double expansion in projectile velocity and $1/M$ in order to identify a simple soft limit.
At the lowest order in $v$ yet to all orders in $1/M$, the differential cross section has a universal form, insensitive to both of the projectile and target spin.
At the relative order-$v^2$, the first two terms in $1/M$ expansion are still independent of the target spin.
The universality starts to be partially violated in the ${\cal O}(v^2/M^2)$ piece, though some terms at this order still remain independent of the target spin, or bears a
definite pattern of target spin dependence.

It is curious that at NNLO in heavy target mass expansion, the prefactors of the $F_{40}=s$ term in both types of gravitational Rutherford scattering processes
encapsulate some peculiar spin-statistics characteristic, which alternate from a constant for fermionic target to another constant for bosonic target.
It is interesting to examine whether this pattern holds true for the composite target with arbitrarily high spin.

\begin{acknowledgments}
We thank Kangyu Chai for participating in the early stage of this work. We are grateful to Zhewen Mo for discussions.
This work is supported in part by the National Natural Science Foundation of China under Grants No. 11925506.
The work of J.-Y. Z. is also supported in part by the US Department of Energy (DOE) Contract No. DE-AC05-06OR23177, under which Jefferson Science Associates, LLC operates Jefferson Lab.
\end{acknowledgments}

\appendix

\section{Polarization sum formula}
\label{appendix:A}

In deriving the unpolarized cross sections, the following spin sum formulas concerning massive target particles are useful:
\begin{subequations}
  \begin{align}
    &\sum_{\lambda}u(p,\lambda)\bar{u}(p,\lambda)=\dfrac{\slashed p+M}{2M},
    \\
    &\sum_{\lambda}\varepsilon_\alpha(p,\lambda)\varepsilon^*_{\alpha'}(p,\lambda)=\hat{\eta}_{\alpha\alpha'},
    \\
    &\sum_{\lambda}u_\alpha(p,\lambda)\bar{u}_{\alpha'}(p,\lambda)=-\dfrac{\slashed p+M}{2M}\left({\eta _{\alpha\alpha'}-\frac{1}{3}\gamma_\alpha\gamma_{\alpha'}-\frac{2p_\alpha p_{\alpha'}}{3M^2}+\frac{\gamma_{\alpha'}p_\alpha-\gamma_{\alpha'}p_{\alpha}}{3M}}\right),
    \\
    &\sum_{\lambda}\varepsilon_{\alpha_1\alpha_2}(p,\lambda)\varepsilon^*_{\alpha'_1\alpha'_2}(p,\lambda)=\hat{\eta}_{\alpha_1\alpha'_1}\hat{\eta}_{\alpha_2\alpha'_2}+\hat{\eta}_{\alpha_1\alpha'_2}
    \hat{\eta}_{\alpha_2\alpha'_1}-\dfrac{2}{3}\hat{\eta}_{\alpha_1\alpha_2}\hat{\eta}_{\alpha'_1\alpha'_2},
\end{align}
\label{eq:spin_sum}
\end{subequations}
with $\hat{\eta}_{\alpha\beta} \equiv -\eta _{\alpha\beta}+\dfrac{p_\alpha p_\beta}{M^2}$. Note that the Dirac
spinor is normalized as $\bar{u}(p,r) u(p,s)=\delta^{rs}$.

\section{The vanishing of the NLO amplitude in the heavy black hole effective theory}
\label{appendix:B}

Since the first detection of the gravitational wave (GW) by LIGO and VIRGO in 2015,
precise predictions of the GW templates becomes an imperative task.
As an efficient theoretical framework to organize the post-Newtonian and post-Minkowski expansion,
the heavy black hole effective theory (HBET) has recently been developed~\cite{Damgaard:2019lfh, Aoude:2020onz},
which is analogous to the heavy quark effective theory (HQET) tailored for heavy quark physics.
Recently it has been applied to GW emission in the scattering of binary spinless black holes or neutron stars with arbitrary masses
at next-to-leading order in the post-Minkowski expansion~\cite{Brandhuber:2023hhy}.

The original HBET Lagrangian is designed to describe a heavy structureless particle (black hole) interacting with soft gravitons,
with the expansion parameter being $1/M$~\cite{Damgaard:2019lfh, Aoude:2020onz}.
We will use this machinery to explain why the NLO amplitude in $1/M$ expansion  vanishes for gravitational Rutherford scattering.

For simplicity, let us consider heavy spinless target particle, which is represented by a complex scalar field.
The underlying theory describing a heavy structureless particle interacting with gravity is simply assumed to
\beq
S = \int d^4 x \sqrt{-g}\left( g^{\mu \nu} \partial_{\mu}\phi^{*} \partial_{\nu}\phi -M^2 | \phi |^2 \right).
\label{Full:Lagrangian}
\eeq

Mimicking the derivation of HQET from QCD, one integrates out the heavy anti-particle degree of freedom by substituting
the equation of motion into \eqref{Full:Lagrangian} and expands the lagrangian in powers of $1/M$.
The effective action in general spacetime background then reads~\cite{Damgaard:2019lfh, Aoude:2020onz}
\begin{align}
    & S_{\rm HBET} = \int d^4x \frac{\sqrt{-g}}{2} \Big[ M\left(v_{\mu} v_{\nu}g^{\mu \nu } -1\right)\varphi_{v}^{*}\varphi_{v} +\frac{i}{2}g^{\mu \nu} \left( v_{\mu} (\varphi_{v}^{*} \partial_{\nu} \varphi_{v} -(\partial_{\nu} \varphi_{v}^{*})\varphi_{v})+v_{\nu} (\varphi_{v}^{*} \partial_{\mu} \varphi_{v} -(\partial_{\mu} \varphi_{v}^{*})\varphi_{v})  \right) \nonumber\\
    & +\frac{i}{2} (v_{\mu} v_{\nu}g^{\mu \nu}-1) \left( (v\cdot \partial \varphi_{v}^{*}) \varphi_{v}-\varphi_{v}^{*}v \cdot \partial \varphi_{v}  \right) \Big].
\label{HBET:Lagrangian}
\end{align}

Since we are interested in the Rutherford scattering in Minkowski spacetime, the weak field approximation $\sqrt{-g}=1-\frac{\kappa}{2}\eta_{\mu \nu} h^{\mu \nu}+\mathcal{O}(h^2)$ is invoked.
The subscript of $\varphi$ is the velocity label of the heavy target particle, $v^\mu=(1,\mathbf{0})$ in the laboratory frame.
It should be cautioned that \eqref{HBET:Lagrangian} was invented for a point-like target particle~\cite{Damgaard:2019lfh, Aoude:2020onz}.
To describe a heavy composite target particle, one should assign general Wilson coefficients $c_i$ to those higher-dimensional operators,
which reflect its nontrivial internal structure.
Keeping the free kinetic term of the heavy scalar field,  as well as organizing the $\varphi_{v}^{*}\varphi_{v}$-graviton interactions in powers of $1/M$,
we then obtain
\begin{align}
    & {\cal L}_{\rm HBET'}=\frac{i}{2}\left( (v\cdot \partial \varphi_{v}^{*}) \varphi_{v}-\varphi_{v}^{*}v \cdot \partial \varphi_{v}  \right)+c_{1}\frac{M\kappa}{2}h^{\mu \nu}v_{\mu}v_{\nu} + \frac{i\kappa }{4} h^{\mu \nu} (c_{2,2} v_{\mu} v_{\nu}+c_{2,1}\eta_{\mu \nu })\left( (v\cdot \partial \varphi_{v}^{*}) \varphi_{v}-\varphi_{v}^{*}v \cdot \partial \varphi_{v}  \right)\nonumber \\
    & +c_{2,1}\frac{i\kappa}{4}h^{\mu \nu} \left( v_{\mu} (\varphi_{v}^{*} \partial_{\nu} \varphi_{v} -(\partial_{\nu} \varphi_{v}^{*})\varphi_{v})+v_{\nu} (\varphi_{v}^{*} \partial_{\mu} \varphi_{v} -(\partial_{\mu} \varphi_{v}^{*})\varphi_{v})  \right),
\label{HBET:Lagrangian:weak field expand}
\end{align}
where the subscript HBET' implies that the original HBET is generalized to account for the composite heavy target particle.

From \eqref{HBET:Lagrangian:weak field expand} one readily reads off the Feynman rules for the $\varphi_{v}^{*}\varphi_{v}$-graviton vertices through NLO in $1/M$:
\begin{subequations}
\bqa
  &&  V^{\varphi_{v}^{*}\varphi_{v} h}_{\text{LO}}=c_{1}\frac{M\kappa}{2}v^{\mu}v^{\nu},
\\
 && V^{\varphi_{v}^{*}\varphi_{v} h}_{\text{NLO}}=\frac{\kappa}{4}\left[ c_{2,1}v^{\mu}(\tilde{p}^{\nu}+\tilde{p}'{}^{\nu})+c_{2,1}v^{\nu}(\tilde{p}^{\mu}+\tilde{p}'{}^{\mu}) -(c_{2,2}v^{\mu} v^{\nu}+c_{2,1}\eta^{\mu \nu }) (v \cdot \tilde{p}+v \cdot \tilde{p}')\right],
\eqa
\label{HBET:three:point:vertices}
\end{subequations}
For a point-like target particle, one simply has $c_{1}=c_{2,1}=c_{2,2}=1$~\footnote{Note that the authors of \cite{Damgaard:2019lfh, Aoude:2020onz} consider the real scalar field.
They simply discard the rapidly oscillating terms proportional to $e^{\pm 2im\cdot v}$ and obtain $c_{1}=c_{2,1}=1$ and $c_{2,2}=0$.}.
Note that the $\tilde{p}$ and $\tilde{p}'$ signify the residual momenta of target particle, {\it e.g.}, $p=Mv+\tilde{p}$.
Assuming the projectile to be a massless spinless point particle, combining \eqref{GFF:point:like:projectile:spin:0}
and the $\varphi_{v}^{*}\varphi_{v} h$ vertices enumerated in \eqref{HBET:three:point:vertices}, we then obtain the HBET prediction to
the gravitational Rutherford scattering amplitudes
through NLO in $1/M$ expansion:
\beq
\mathcal{M}_{\text{EFT}}=\frac{c_{1} \kappa^2 M}{4(\cos{\theta}-1)}+\frac{c_{2,2}\kappa^2(|{\bf k}|-|{\bf k}'|
)}{16 \sin^2 \left(\frac{\theta}{2}\right)}+\mathcal{O}\left(\frac{1}{M}\right),
\label{EFT:amplitude}
\eeq
However, in light of the relation between $|{\bf k}|$ and $|{\bf k}'|$ as given in \eqref{kprime:k:relation}, 
one readily observes that the second (nominally NLO) term in \eqref{EFT:amplitude} is suppressed with respect to the LO term 
actually by a factor of $1/M^2$, rather than $1/M$. Therefore, we conclude
\beq
\mathcal{M}_{\text{EFT}}=\frac{c_{1} \kappa^2 M}{4(\cos{\theta}-1)}+\mathcal{O}\left(\frac{1}{M}\right).
\label{EFT:amplitude:up:to:NLO}
\eeq
This EFT analysis provides a clear perspective to understand why the NLO amplitude vanishes.

The EFT prediction \eqref{EFT:amplitude:up:to:NLO} should be identical to the LO result in 
\eqref{cross:section:Rutherford:massless:spin:0:projectile:LO}, which has been derived earlier 
in terms of the GFFs of composite target particle,
\beq
\mathcal{M}=\frac{\kappa^2 M^2 F_{10}}{2(\cos{\theta}-1)}+ \mathcal{O}\left(\frac{1}{M^0}\right).
\eeq
This criterion enforces $c_{1}=F_{10}=1$, once the $2M$ factor is compensated for the non-relativistic state normalization. 
The requirement $c_1=1$, irrespective to whether the heavy target particle is fundamental or composite, may be attributed to the 
reparametrization invariance in HBET.

Since the ${\cal O}(1/M)$ HBET vertex yields a vanishing contribution to the Rutherford scattering amplitude, therefore the NLO contribution to the 
unpolarized cross section in \eqref{cross:section:Rutherford:massless:spin:0:projectile:NLO} 
solely arises from the expansion of the phase space factor.

\end{document}